\documentclass[12pt]{article}  
\usepackage{epsfig}
\def\lapproxeq{\lower .7ex\hbox{$\;\stackrel{\textstyle <}{\sim}\;$}}
\def\gapproxeq{\lower .7ex\hbox{$\;\stackrel{\textstyle >}{\sim}\;$}}
\def\Im{\mathop{\rm Im}\nolimits}

\begin{document} \begin{titlepage}
\begin{flushright}
August, 2005\\
MSUHEP-050715\\
hep-ph/0508311
\end{flushright}
\vspace*{0.2cm}
\begin{center} \begin{large}
{\bf Enhancement of ``CP-odd'' Higgs Boson Production\\
in the Minimal Supersymmetric Standard Model\\
with Explicit CP Violation}
\end{large} \end{center}

\begin{center}
\vspace*{0.2cm} 
Qing-Hong Cao, Daisuke Nomura, Kazuhiro Tobe, and C.-P. Yuan\\
\vspace*{0.2cm}
Department of Physics and Astronomy\\
Michigan State University\\
East Lansing, MI 48824, USA\\
\end{center}

\vspace*{1cm}

\begin{abstract}
\noindent
We calculate the production cross section of the ``CP-odd''
Higgs boson via gluon fusion in the minimal supersymmetric standard
model with explicit CP violation in the stop sector.  
We show that there is a parameter region in which the cross section 
is enhanced by a factor of about 1000, as compared to the case
without CP violation in the stop sector.  In the parameter region
where the ``CP-odd'' Higgs boson can decay into a stop pair, the stop
pair events will be the important signature of
the enhanced ``CP-odd'' Higgs boson. In the case where the 
``CP-odd'' Higgs boson cannot decay into any superparticles,
the $\gamma\gamma$ and $\tau\tau$ decay channels
could become important for discovering
the ``CP-odd'' Higgs boson.  We also discuss the constraints from 
electric dipole moments of electron, neutron and mercury on the viable
parameter space mentioned above.
\end{abstract}
\end{titlepage}

\newpage

Low energy supersymmetry (SUSY) is one of the most promising
candidates of physics beyond the Standard Model (SM).  SUSY
gives an elegant solution to the naturalness problem of the
stability of the weak scale by canceling quadratically divergent
radiative corrections.

One of the most important predictions of the minimal supersymmetric
standard model (MSSM) is the upper bound of the lightest Higgs boson mass.
At tree level, the MSSM predicts the lightest Higgs boson mass to be
less than the $Z$ boson mass.  However, after including
loop corrections, the contributions from top and stop loops
are so important that the upper bound of the lightest Higgs boson
mass can be increased to around 130 GeV~\cite{LightestHiggsMass}.
This upper bound should be compared with the current lower
limit of 89.8 GeV from the MSSM Higgs search at
LEP~\cite{Eidelman:2004wy}. 
If the lightest Higgs boson is discovered and its mass turns out to be
less than 130 GeV, it is a strong hint for the MSSM.

If the MSSM is true, the CERN Large Hadron Collider (LHC) 
is expected to probe the Higgs 
sector by copiously producing the Higgs bosons.  The Higgs sector 
in the MSSM has a rich structure; there are two CP-even
Higgs bosons, one CP-odd Higgs boson and one (complex) charged Higgs boson.
Their production and decay properties depend on various parameters in the
MSSM including the SUSY breaking parameters. 
Therefore, to study the properties of the Higgs bosons at the LHC, 
a precise knowledge of the production cross section of the Higgs bosons
is extremely important. 

It has been shown that CP-violation in the Higgs sector
could significantly affect the production and decay properties of
the Higgs bosons~\cite{CP_Higgs,Choi:2001iu,Dedes:1999sj}. 
In order to prepare for 
the discoveries of the MSSM Higgs bosons at the LHC in any case, further 
detailed studies on the MSSM with CP-violation would be important.  
The aim of this letter is to present our findings on
the production cross section of the ``CP-odd'' Higgs boson in the
MSSM with CP-violation.\footnote{
Strictly speaking, when CP is violated,
we cannot define a ``CP-odd'' Higgs boson because all three neutral 
Higgs bosons 
are mixed with each other. As we will discuss later, however, 
in the parameter sets we consider, CP-violating
Higgs boson mixing is small.
Therefore, we still use the
terminology ``CP-odd'' Higgs boson even in the CP-violating case.}
We show that the production cross section of the ``CP-odd'' Higgs 
boson can be enhanced by a factor of about 1000 compared to
the case without CP-violation, and discuss some important decay
signatures of the ``CP-odd'' Higgs boson.\footnote{
Although in this letter we concentrate on the ``CP-odd'' Higgs production via
gluon fusion at hadron colliders, we note that the same enhancement of the
``CP-odd'' Higgs production is also possible at a $\gamma \gamma$ collider.
}
  We also discuss some constraints
on our CP-violating scenarios.
The strongest constraint comes from the electric dipole moments (EDMs)
of electron and neutron.
Since there are possibilities that cancellations among many
contributions to EDMs could happen, the searches for the
``CP-odd'' Higgs boson at the current and future colliders could provide
important information on the CP-violation mechanism in the MSSM, which
is generally independent of those from the EDM searches.

The MSSM has two Higgs doublets, $H_1$ and $H_2$. 
The neutral components $H_1^0$ and $H_2^0$ of the Higgs bosons
develop vacuum expectation values (VEVs), which trigger the electroweak 
symmetry breaking (EWSB).  After EWSB, there are three neutral 
Higgs bosons and a pair of charged Higgs bosons.  If CP is a good 
symmetry in the Higgs sector, 
we can label the neutral Higgs bosons in terms of CP properties as two 
CP-even Higgs bosons $h^0$ and $H^0$, and a CP-odd Higgs boson $A$.  
In general, 
if CP is violated in the sfermion sector, CP-violating mixing 
among the three Higgs bosons is induced through radiative corrections.  
In this letter, we consider the CP violation in the Higgs
sector radiatively induced by the trilinear coupling of stop 
$A_t$,\footnote{The complex trilinear coupling of sbottom $A_b$
could also induce an important effect similar to the one discussed 
in this letter.  For simplicity, however, we assume $A_b$ to be a real 
parameter.}
which is defined as
\begin{eqnarray}
 {\cal L} = 
- \left( \frac{\sqrt2 m_t}{v \sin\beta} A_t H_2 \tilde{t}_R^\ast  \tilde{q}_L
+ {\rm h.c.} \right),
\label{eq:stop_Aterm}
\end{eqnarray}
where $H_2$ is the Higgs doublet that generates
top quark mass $m_t$ via Yukawa interaction, 
$\tilde{q}_L$ is the third generation squark doublet,
and  $\tilde{t}_R$ is the right-handed stop. 
In our notation, $\phi_1$ and $\phi_2$ ($a_1$ and $a_2$) are the real 
(imaginary) components of $H_1^0$ and $e^{-i\xi} H_2^0$, respectively, 
which are explicitly given by
\begin{eqnarray}
 H_1^0 = \frac1{\sqrt2} \left(\phi_1 + v_1+i a_1 \right), ~~~~~
 H_2^0 = \frac{e^{i\xi}}{\sqrt2} \left(\phi_2 + v_2+i a_2 \right).
\end{eqnarray}
The VEV $v_1$ is relevant to the masses of down-type 
quarks and leptons, and $v_2$ is responsible for the up-type quark
masses.  The ratio of the two VEVs is parametrized by 
$\tan\beta \equiv v_2/v_1$, and $v$ is defined as 
$v \equiv \sqrt{v_1^2+v_2^2}$, which is about 246 GeV.
In general the relative phase $\xi$ of the VEVs can be 
non-zero.  For simplicity, in this letter we do not consider 
the effect of non-vanishing $\xi$ and set $\xi=0$ in the following.   
One of the linear combinations $(G)$ of the CP-odd components 
$a_1$ and $a_2$ is eaten by the $Z$ boson $(G=a_1\cos\beta-a_2\sin\beta)$, 
and the other linear combination $(A)$
becomes the physical ``CP-odd'' Higgs boson $(A=a_1\sin\beta+a_2\cos\beta)$.
Once we allow the $A_t$ parameter to be complex, it induces
CP-violating mixing among the neutral Higgs bosons.
The CP-violating elements of the mass-squared matrix ${\cal M}^2_H$ 
at one-loop level are given as
\begin{eqnarray}
{\cal M}^2_H \bigg|_{A \phi_1} 
= \frac{3}{16\pi^2} 
  \frac{m_t^2}{\sin\beta}
\frac{\Im(A_t \mu)}{m^2_{\tilde{t}_2}-m^2_{\tilde{t}_1}} F_{t},~~~~~~~~
{\cal M}^2_H \bigg|_{A \phi_2} 
=  \frac{3}{16\pi^2} 
  \frac{m_t^2 }{\sin\beta}
\frac{\Im(A_t \mu)}{m^2_{\tilde{t}_2}-m^2_{\tilde{t}_1}} G_{t},
\end{eqnarray}
where the explicit forms of the dimensionless quantities 
$F_t$ and $G_t$ were given in Ref.~\cite{Choi:2000wz}.  In the
equations above, ${\cal M}^2_H \big|_{A \phi_{1(2)}}$ is
the $(A,\phi_{1(2)})$ element of the mass-squared matrix ${\cal M}^2_H$. 
$m_{\tilde{t}_1}$ and $m_{\tilde{t}_2}$ are the lighter and the 
heavier stop masses, respectively.  
In general, the Higgsino mass parameter $\mu$ as well as
$A_t$ can have a CP violating phase.  
For simplicity, we assume that only the trilinear coupling $A_t$ 
is complex and  $\mu$ is real.
Because of the mixing induced by the CP-violating coupling $A_t$, 
mass eigenstates of neutral Higgs bosons
$(h_1, h_2, h_3)$ are linear combinations of 
the three neutral Higgs bosons $\phi_1$, $\phi_2$ and $A$:
\begin{eqnarray}
\left( \begin{array}{c} h_1\\ h_2 \\ h_3 \end{array} \right)_i
= 
O_{i \alpha}
\left( \begin{array}{c} \phi_1\\ \phi_2 \\A \end{array} \right)_\alpha~,
\end{eqnarray}
where $O_{i \alpha}$ is the orthogonal matrix which diagonalizes
${\cal M}^2_H$, and the label of the mass eigenstates is determined 
in such a way that the masses $m_{h_1}, m_{h_2}, m_{h_3}$ satisfy
$m_{h_1} \leq m_{h_2} \leq m_{h_3}$.  It has been pointed 
out~\cite{CP_Higgs,Choi:2001iu,Dedes:1999sj}
that in some parameter regions the induced mixing can be large and
play an important role in Higgs physics.
However, in this letter, we focus on the regions of the SUSY
parameter space in which the mixing with ``CP-odd'' Higgs boson is
small and the second lightest Higgs boson $h_2$ is almost a ``CP-odd'' 
Higgs boson (typically $|O_{23}|^2>0.9$). Therefore, in the qualitative 
discussion 
below, we neglect the mixing effects and we still use the terminology 
``CP-odd'' Higgs boson. However, in our numerical results to be
shown below, we include the mixing 
effects, and we call the second lightest Higgs boson $h_2$ the ``CP-odd'' 
Higgs boson $A$.

Now we are ready to discuss the Higgs boson production cross section.
For the lightest Higgs boson $h^0(=h_1)$, it is known that the radiatively
induced CP-violation can significantly change the cross section of
$gg\to h^0$~\cite{Choi:2001iu,Dedes:1999sj}. 
In this letter we consider the production of the ``CP-odd''
Higgs boson $A$.\footnote{
Similar analyses had been done in 
Refs.~\cite{Dedes:1999sj,A_production_in_CPVMSSM}.  
The authors of those 
articles performed the analyses for the parameter sets
different from those discussed here.}
This is motivated by the following reason.

\begin{figure}
\centering
\includegraphics*[width=16cm,angle=0]{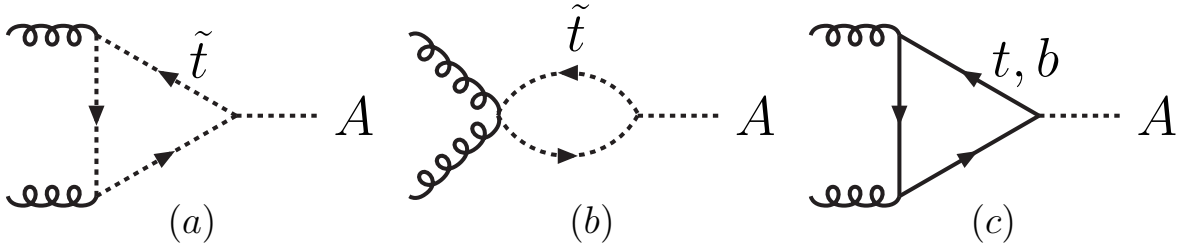}
\caption{The Feynman diagrams which contribute to
$gg \to A$ in the MSSM with CP-violation when CP violating 
mixing among Higgs bosons are neglected.  If the trilinear
coupling $A_t$ is complex, there is a finite contribution 
from the diagrams $(a)$ and $(b)$ to the total production cross section.
If there is no CP-violations in the sfermion sector, 
the diagrams $(a)$ and $(b)$ do not contribute to the total cross section.
The contribution from the diagram $(c)$ is always there,
even in the CP-conserving case.}
\label{fig:stoploop}
\end{figure}

If CP is not violated, the most important contribution to 
$gg\to A$ comes from the diagram $(c)$ in Fig.~\ref{fig:stoploop}.
In the language of effective Lagrangian, this diagram is described
by the CP-even operator,
\begin{eqnarray}
 {\cal L} = c^A_{t/b} A G^{a\mu\nu} \widetilde{G}^a_{\mu\nu} ,
\label{eq:top_contr}
\end{eqnarray}
where the coefficient $c^A_{t/b}$ is obtained by integrating out 
the top and the bottom loops.  $G^a_{\mu\nu}$ is the field
strength tensor for gluon with $a$ being a color index 
($a=1, \ldots, 8$), and $\widetilde{G}^a_{\mu\nu}$ is its
dual, $\widetilde{G}^a_{\mu\nu} \equiv 
\epsilon_{\mu\nu\rho\sigma} G^{a \rho\sigma}/2$.
Note that the stop diagrams shown in Fig.~\ref{fig:stoploop} $(a)$
and $(b)$ do not contribute to $gg\to A$ simply because the couplings 
of the $\tilde{t}^*_i \tilde{t}_i A~(i=1,2)$ interactions  
vanish due to the CP symmetry.\footnote{In other words,
this can be understood by the cancellation between diagrams of
left- and right-handed stop loop contributions 
in the weak eigenstate basis.}
Therefore, the leading order (LO) parton-level cross section of 
$gg\to A$ in the CP-conserving (CPC) case, 
$\sigma^{\rm LO}(gg\to A)_{\rm CPC}$, is given by the 
top/bottom contributions alone: 
\begin{eqnarray}
 \sigma^{\rm LO}(gg \to A)_{\rm CPC} \propto  \left| c^A_{t/b} \right|^2 
\label{eq:ggtoA_top}
\end{eqnarray}

On the other hand, in the CP-violating (CPV) case, the couplings
$\tilde{t}^*_i \tilde{t}_i A~(i=1,2)$ are not zero. Hence, the stop
diagram contributes to the Higgs boson production $gg\to A$.  An important 
point is that the effective operator induced by the diagrams 
$(a)$ and $(b)$ of Fig.~\ref{fig:stoploop} is CP-odd, 
\begin{eqnarray}
 {\cal L} = c^A_{\tilde{t}} A G^{a\mu\nu} G^a_{\mu\nu} ,
\label{eq:stop_contr}
\end{eqnarray}
where the coefficient $c^A_{\tilde{t}}$ is
determined from the stop loop contribution. 
Since the CP-properties of the operators in
Eqs.~(\ref{eq:top_contr}) and (\ref{eq:stop_contr})
are opposite and the total cross section 
is a CP-even quantity,
these two contributions do not interfere with each other 
in the total cross section. Hence, the LO total cross section 
$\sigma^{\rm LO}(gg\to A)_{\rm CPV}$ in the CP-violating case 
is proportional to the sum of 
the squares of the contributions from these diagrams:
\begin{eqnarray}
 \sigma^{\rm LO}(gg \to A)_{\rm CPV} \propto \left( 
        \left| c^A_{t/b} \right|^2 
      + \left| c^A_{\tilde{t}} \right|^2 \right) .
\label{eq:ggtoA}
\end{eqnarray}
Note that in the case of the CP-even Higgs boson
production, both the top/bottom and the stop/sbottom
loops contribute to $gg\to h$ even when CP is conserved, and
generate the same effective operator,
\begin{eqnarray}
 {\cal L} = (c^h_{\tilde{t}/\tilde{b}} + c^h_{t/b})
            h G^{a\mu\nu} G^a_{\mu\nu} , 
\label{eq:GG}
\end{eqnarray}
so that they could interfere with each other. Here, $h$ represents
the ``CP-even'' Higgs bosons, $h^0$ and $H^0$.
When CP is violated, the induced operator is the same
as the one in Eq.~(\ref{eq:GG}) (with a different coefficient)
at the leading order, and the interference indeed
can significantly affect the production cross 
section~\cite{Choi:2001iu,Dedes:1999sj}.
Therefore the effect of CP-violation on the ``CP-odd'' Higgs
boson production
is quite different from that on the ``CP-even'' Higgs bosons, and 
the cross section
of ``CP-odd'' Higgs boson in the CP-violating case is always enhanced 
by the stop contribution, compared 
to the one in the CP-conserving case.
Thus, it is interesting to study the ``CP-odd'' Higgs boson production in 
the CP-violating case in order 
to see how large enhancement can be induced by the CP-violating 
interaction $A_t$.

\begin{figure}[ht]
\centering
\includegraphics*[width=9.0cm,angle=0]{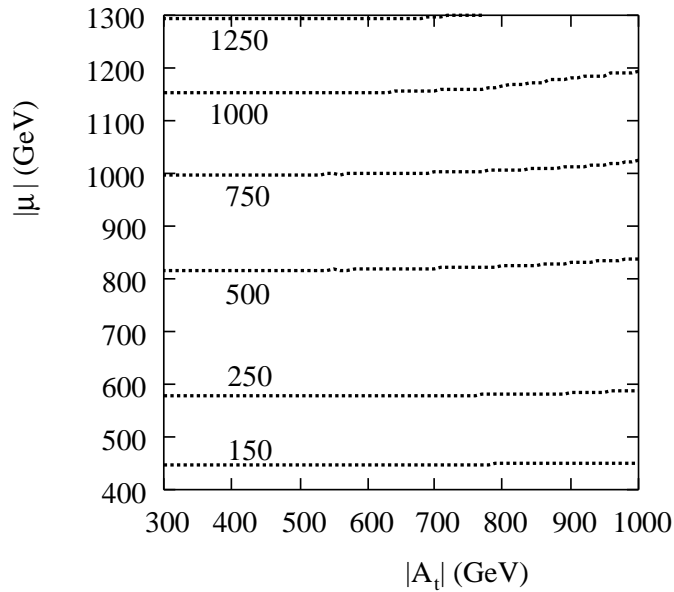}
\caption{The contour plot of the
ratio of the LO parton-level cross sections
in the CP-violating (CPV) case and the CP-conserving (CPC) case 
$\sigma^{\rm LO}(gg\to A)_{\rm CPV}/\sigma^{\rm LO}(gg\to A)_{\rm CPC}$
as a function of $|A_t|$ and $\mu$.  The SUSY parameters are
fixed as in Eq.~(\ref{eq:sample}).}
\label{fig:crosssection}
\end{figure}

Our numerical results on the ratio 
$\sigma^{\rm LO}(gg \to A)_{\rm CPV}/
\sigma^{\rm LO}(gg\to A)_{\rm CPC}$
are shown as a function of $|A_t|$ and $\mu$ 
in Fig.~\ref{fig:crosssection}.
In the figure we have taken the sample parameter set as,
\begin{eqnarray}
 m_A = 250~{\rm GeV}, ~~~~ m_{\tilde{t}_1} = 120~{\rm GeV},
~~~~~ \tan\beta=6,~~~~~
  {m}_{\tilde{t}_L} =  {m}_{\tilde{t}_R},~~~~~A_t=i|A_t|,
~~~~~\mu=|\mu|,
\label{eq:sample}
\end{eqnarray}
where ${m}_{\tilde{t}_L} ({m}_{\tilde{t}_R})$
is the soft SUSY breaking mass for the left-handed (right-handed) stop.
We see that the cross section can be enhanced by a factor
of about 1000, compared to the case without CP violation.
This huge enhancement can be understood in the following way.
If we neglect the CP-violating mixing among Higgs bosons,
the ratio 
$\sigma^{\rm LO}(gg\to A)_{\rm CPV}/\sigma^{\rm LO}(gg\to A)_{\rm CPC}$ 
can be written as
\begin{eqnarray}
\frac{\sigma^{\rm LO}(gg\to A)_{\rm CPV}}
{\sigma^{\rm LO}(gg\to A)_{\rm CPC}}
=\left| 
\frac{c^A_{\tilde{t}}}{c^A_{t/b}} 
\right|^2 + 1,
\end{eqnarray}
for the same $m_A$ and $\tan\beta$ in both cases.
Explicitly calculating the top/bottom loop and the stop loop 
diagrams, we obtain
\begin{eqnarray}
\frac{\sigma^{\rm LO}(gg\to A)_{\rm CPV}}
{\sigma^{\rm LO}(gg\to A)_{\rm CPC}}
=
\frac{m_t^2}{m_A^4} \frac{|\mu A_t|^2(1 + \cot^2\beta)^2}
{|A_t|^2+|\mu \cot\beta|^2}\frac{|m^2_{\tilde{t}_1}
C_0(m^2_{\tilde{t}_1},m_A^2)-m^2_{\tilde{t}_2}
C_0(m^2_{\tilde{t}_2},m_A^2)|^2}
{|m_t^2 C_0(m^2_{t},m_A^2)\cot\beta
+m_b^2 C_0(m^2_{b},m_A^2)\tan\beta|^2} + 1,
\label{eq:ratio_exp}
\end{eqnarray}
where, for simplicity, we have assumed that $A_t$ is pure imaginary,
$\mu$ is real, and the mixing between stops is maximal, {\it i.e.},
$m^2_{\tilde{t}_{LL}}=m^2_{\tilde{t}_{RR}}$ where 
$m^2_{\tilde{t}_{LL}}$ and $m^2_{\tilde{t}_{RR}}$ are
$(\tilde{t}_L,\tilde{t}_L)$ and $(\tilde{t}_R,\tilde{t}_R)$ 
elements of the stop mass matrix, respectively.
The function $C_0$ is a one-loop function~\cite{'tHooft:1978xw}.
For our particular case here, we define it as
\begin{eqnarray}
 C_0(m^2,m_A^2) = \frac1{i\pi^2} 
 \int \frac{d^4 q}{(q^2-m^2)((q+p_1)^2-m^2)((q+p_1+p_2)^2-m^2)},
\end{eqnarray}
where $p_1^2=p_2^2=0$ and $(p_1+p_2)^2=m_A^2$.
If $m_A < 2 m_{\tilde{t}_1}$, 
$|m_{\tilde{t}_1}^2 C_0(m_{\tilde{t}_1}^2,m_A^2)
- m_{\tilde{t}_2}^2 C_0(m_{\tilde{t}_2}^2,m_A^2)|^2$ 
term in Eq.~(\ref{eq:ratio_exp}) is 
the square of a subtraction of a real number from another real 
number, where a GIM-like cancellation happens.
When $2 m_{\tilde{t}_1}< m_A <2 m_{\tilde{t}_2}$, 
which is satisfied for our sample parameters,
the function $C_0(m_{\tilde{t}_1}^2,m_A^2)$ develops an imaginary part
(when crossing the mass threshold for producing a light stop pair)
and the factor is a subtraction of a real number from a complex
number, which means the cancellation tends to be less severe.  
Since in our sample parameter set $m_A< 2 m_{t}$, 
$C_0(m_t^2,m_A^2)$ in the denominator does not have an imaginary part,
which also makes the ratio larger.
(For moderate $\tan\beta$, the $C_0(m_b^2,m_A^2)$ term is not
very important.)
In addition, when $|A_t| \gg \mu\cot\beta$, the ratio in 
Eq.~(\ref{eq:ratio_exp}) behaves like $|\mu|^2$,
as can be seen in Fig.~\ref{fig:crosssection}.
Therefore large $|A_t|$ and $\mu$ also induce large 
enhancement in the ratio.\footnote{A large $|A_t|$ may be dangerous
because it could develop a color breaking VEV~\cite{color_breaking}. 
Here, we have checked that
the large part of our parameter space ($|A_t| \lapproxeq 950$ GeV)
satisfies the condition 
$|A_t|^2 
< 3(m^2_{\tilde{t}_L}+m^2_{\tilde{t}_R}+m_{H_2}^2+|\mu|^2)$, 
which guarantees to avoid a color breaking VEV in a $D$-flat direction
$|\tilde{t}_L|=|\tilde{t}_R|=|H_2^0|$ at the tree level potential.}

\begin{figure}
\centering
\includegraphics*[width=10cm,angle=0]{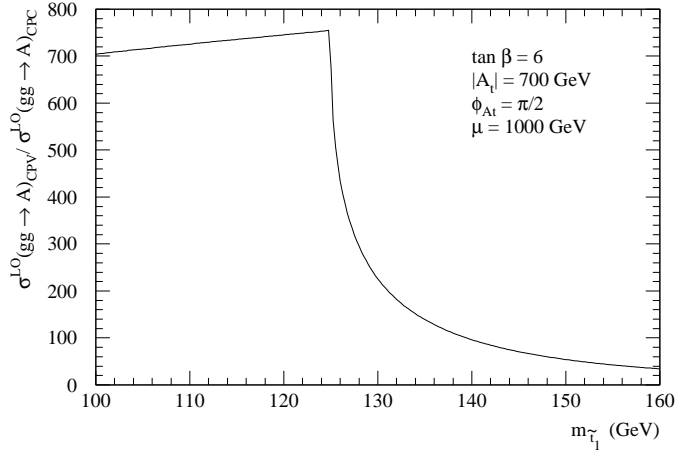}
\caption{The ratio of the LO parton-level cross sections
in the CP-violating (CPV) case and the CP-conserving (CPC) case, 
$\sigma^{\rm LO}(gg\to A)_{\rm CPV}/
\sigma^{\rm LO}(gg\to A)_{\rm CPC}$,
as a function of $m_{\tilde{t}_1}$.  
Here we took $\tan\beta=6$, $m_A=250$ GeV, $|A_t|=700$ GeV,
$\phi_{A_t}=\pi/2$ and $\mu=1$ TeV.
The LO hadron-level cross sections of the ``CP-odd'' Higgs boson via
gluon fusion in the CP-conserving case
are $0.8$ fb and $0.2$ pb at the Tevatron
and the LHC, respectively, for $m_A=250$ GeV and 
$\tan\beta=6$.
}
\label{fig:ratio_stop}
\end{figure}

In Eq.~(\ref{eq:ratio_exp}), we have not included the effect
from the mixing among the Higgs bosons although we have included
that effect in the numerical results shown in Fig.~\ref{fig:crosssection}.  
We have checked that the second lightest Higgs boson $h_2$
is almost a ``CP-odd'' Higgs boson for our sample 
parameter sets.  In fact, 
$|O_{23}|^2>0.9$
for $2.3 |A_t| - \mu \gapproxeq  100~{\rm GeV}$, 
and $|O_{23}|^2>0.7$
for $5 |A_t| - \mu \gapproxeq 350~{\rm GeV}$
in the range shown in the figure.

In Fig.~\ref{fig:ratio_stop}, we also show the ratio
$\sigma^{\rm LO}(gg\to A)_{\rm CPV}/\sigma^{\rm LO}(gg\to A)_{\rm CPC}$ 
as a function of
$m_{\tilde{t}_1}$ while fixing $m_A$ and $\tan\beta$. Here, we took 
the same sample parameters as given in Eq.~(\ref{eq:sample}) except that
we set $|A_t|$ and $\mu$ to be $700$ GeV and $1$ TeV, respectively.
As can be seen from  Fig.~\ref{fig:ratio_stop}, as $m_{\tilde{t}_1}$ gets 
larger than $m_A/2$, the ratio rapidly drops off because of the 
GIM-like cancellation in the 
$|m_{\tilde{t}_1}^2 C_0(m_{\tilde{t}_1}^2,m_A^2)
- m_{\tilde{t}_2}^2 C_0(m_{\tilde{t}_2}^2,m_A^2)|^2$ 
term in Eq.~(\ref{eq:ratio_exp}). 
However, due to the enhancement by large $|A_t|$ 
and $\mu$, the ratio can still be of ${\cal O}(100)$ if the stop mass
is near the threshold $m_{\tilde{t}_1}\sim m_A/2$.

\begin{table}
\begin{tabular}{|c||c|c|c|c|}
\hline
Tevatron ($\sqrt{s}=1.96$ TeV) & $\sigma^{\rm LO}(A)$  
& BR($A\to \tilde{t}_1^* \tilde{t}_1$) & BR($A\to \gamma \gamma$)&
BR($A\to \tau \tau$) \\
\hline
CPC case & $0.8$ fb & 0 & $\sim 10^{-4}$ & $\sim 0.05$ \\ 
\hline
CPV case ($m_{\tilde{t}_1}=120$ GeV) & $\sim 110-1200$ fb 
    & $\sim 1$ & ${\cal O}(10^{-5})$ & ${\cal O}(10^{-3})$ \\
\hline\hline
LHC ($\sqrt{s}=14$ TeV) & $\sigma^{\rm LO}(A)$ 
& BR($A\to \tilde{t}_1^* \tilde{t}_1$) & 
BR($A\to \gamma \gamma$)& BR($A\to \tau \tau$)\\
\hline
CPC case & $0.2$ pb & 0 & $\sim 10^{-4}$ & $\sim 0.05$\\ 
\hline
CPV case ($m_{\tilde{t}_1}=120$ GeV) & $\sim 30-300$ pb & $\sim 1$ 
 & ${\cal O}(10^{-5})$ & ${\cal O}(10^{-3})$\\
CPV case ($m_{\tilde{t}_1}=130$ GeV) & $\sim 10-90$ pb & 0 
& ${\cal O}(10^{-4})$ & ${\cal O}(10^{-1})$ \\
\hline
\end{tabular}
\caption{The leading order (LO) hadron-level cross 
sections of the ``CP-odd'' Higgs boson production via gluon 
fusion ($\sigma^{\rm LO}(A)$) at the Tevatron ($\sqrt{s}=1.96$ TeV) 
and the LHC ($\sqrt{s}=14$ TeV)
and the decay branching ratios of $A$ into 
$\tilde{t}_1^*\tilde{t}_1$, $\gamma\gamma$,
$\tau\tau$ are shown in the CP-conserving (CPC) case 
and the CP-violating (CPV) case discussed in this letter.
Here for the CPV case we took $m_A=250$ GeV, $\tan\beta=6$, 
$400$ GeV $< \mu<1300$ GeV and $300$ GeV $<|A_t|<1000$ GeV.
For the calculation of the branching ratios in the CPC case 
we took $m_A=250$ GeV, $m_{\tilde{t}_1}=130$ GeV, $\tan\beta=6$, 
$\mu=700$ GeV and $A_t=700$ GeV as an example.}
\label{table:summary}
\end{table}

In Table~\ref{table:summary}, we summarize our results.
In the table, we list the LO hadronic-level 
cross sections of the ``CP-odd'' Higgs boson $A$
via gluon fusion ($\sigma^{\rm LO}(A)$) 
at the Tevatron ($\sqrt{s}=1.96$ TeV)
and the LHC ($\sqrt{s}=14$ TeV),
the branching ratios BR($A\to \tilde{t}_1^*\tilde{t}_1$), 
BR($A\to \gamma\gamma$), and BR($A\to \tau\tau$) in 
various cases discussed in this letter.
The LO cross sections are calculated using the CTEQ6L parton 
distribution functions~\cite{Pumplin:2002vw}\footnote{
The QCD corrections to the production cross section 
of the CP-odd Higgs boson are known up to and including the 
next-to-next-to-leading order (NNLO) in the CP-conserving 
MSSM~\cite{NNLO}. 
When we parametrize the hadron-level higher 
order (HO) production cross section $\sigma^{\rm HO}(pp \to A)$
of the CP-odd Higgs boson using the LO hadron-level cross section
$\sigma^{\rm LO}(pp \to A)$
as $\sigma^{\rm HO}(pp \to A)=K \sigma^{\rm LO}(pp \to A)$, 
the $K$ factor is found to be approximately 2 for $m_A=250$ GeV 
and $\sqrt{s}=14$ TeV in the CP-conserving MSSM at NNLO QCD~\cite{NNLO}. 
In the CP-violating case we expect the $K$ factor to be almost 
the same as in the CP-conserving case, which is,
however, beyond the scope of this letter.}, 
and the branching ratios of the ``CP-odd'' Higgs boson $A$ are 
computed using a publicly available code ``CPsuperH''~\cite{CPsuperH}.

The LO cross sections of the ``CP-odd'' Higgs boson via
gluon fusion in the CP-conserving case
are $0.8$ fb and $0.2$ pb at the Tevatron
and the LHC, respectively, for $m_A=250$ GeV and 
$\tan\beta=6$.  
These cross sections are not large enough to allow us
to discover the CP-odd Higgs boson at the 5$\sigma$ level
even at the LHC~\cite{ATLAS_TDR}.
On the other hand, in the CP-violating case with 
$m_A=250$ GeV, $\tan\beta=6$, and $m_{\tilde{t}_1}=120$ GeV, 
we can read from Fig.~\ref{fig:crosssection}
that the LO cross section can be as large as $110-1200$ fb at the 
Tevatron, and $30-300$ pb at the LHC 
for $400$ GeV $< \mu<1300$ GeV and $300$ GeV $<|A_t|<1000$ GeV.
In the CP-violating case with $m_A=250$ GeV 
and $m_{\tilde{t}_1}=120$ GeV, the ``CP-odd'' Higgs boson can decay into 
a stop pair. Since the coupling of the ``CP-odd'' Higgs boson to stops 
is large, we found that the branching ratio 
BR($A\to \tilde{t}_1^* \tilde{t}_1$) is almost one. Therefore, 
the stop pair production via the ``CP-odd'' Higgs boson production can be 
one of important signatures of the ``CP-odd'' Higgs boson in the
CP-violating case.  At the Tevatron, 
$\sigma\times {\rm BR}(A\to \tilde{t}_1^* \tilde{t}_1)$
can be $\sim 110-1200$ fb in the LO calculation. 
This stop production cross section via $A$-decay
is smaller than the normal stop production
cross section which is about 10 pb~\cite{stop_search}.  At the LHC,
$\sigma\times {\rm BR}(A\to \tilde{t}_1^* \tilde{t}_1)$ can be as large as
$\sim 30-300$ pb. Thus, it might be possible to detect the ``CP-odd''
Higgs boson $A$ in the stop pair channel, 
although a detailed study for this process is needed.
When $m_A < 2 m_{\tilde{t}_1}$, the ``CP-odd'' Higgs boson is not 
kinematically allowed to decay into a stop pair 
(and into any SUSY particle pairs if $2m_{\rm LSP}>m_A$, where 
$m_{\rm LSP}$ is the lightest superparticle mass),
though the production cross section of $A$ can still be large.
For example, in the case with $m_A=250$ GeV and $m_{\tilde{t}_1}=130$ GeV
the LO cross section is about $\sim 10-90$ pb.
As shown in Table~\ref{table:summary}, 
$\sigma\times {\rm BR}(A\to \gamma \gamma)$
can be ${\cal O}(10)$ fb at the LHC in the leading order calculation. 
Comparing this result with the one analyzed in the ATLAS TDR~\cite{ATLAS_TDR}, 
the LHC with an integrated luminosity of 100 fb$^{-1}$ or more may be 
able to discover the ``CP-odd'' Higgs boson $A$ via the diphoton mode.
Also the $A\to \tau \tau$ mode would be important, for
its decay branching ratio is much larger than the diphoton mode.  From
Table~\ref{table:summary}, $\sigma \times {\rm BR}(A \to\tau\tau)$ 
can be ${\cal O}(10)$ pb which is large enough to be detected at the 
LHC~\cite{ATLAS_TDR,Belyaev:2005ct}.  
Although the branching ratio of $A \to \mu \mu$ is suppressed by a factor
of $(m_\mu/m_\tau)^2$ compared to the branching ratio of $A \to \tau \tau$,
the $A \to \mu \mu$ channel could also be useful for studying
the ``CP-odd'' Higgs boson in some parameter regions.
The branching ratio of $A\to Zh$ is 
not large (at most 1-2 \% for our parameter sets).  This can be
understood by the fact that in the 
decoupling limit $m_A \gg m_Z$, ${\rm BR}(A\to Zh)$ is zero in the
CP-conserving case, and for the parameter sets studied
in this letter in the CP-violating case, the ``CP-odd'' Higgs 
boson is heavy enough that the decoupling limit also holds.
In summary, in the presence of CP-violation in the Higgs sector, 
the discovery potential
for the ``CP-odd'' Higgs boson at the Tevatron and the LHC could be 
strongly modified.

Finally we would like to discuss some constraints on the CP-violating 
cases discussed in this letter.
The first one is the lightest Higgs boson mass bound. Since in our 
CP-violating scenarios the heavier Higgs bosons are heavy enough so that
the coupling of the $ZZh$ interaction is not very different from 
the one in the SM, the lower limit on the SM Higgs boson mass 
$m_h>114$ GeV would still apply.
Using ``CPsuperH''~\cite{CPsuperH}, we have checked the 
lightest Higgs boson mass limit is satisfied for $500$ GeV$<|A_t|<900$ GeV.
The second constraint is from electroweak precision measurement. 
Since the stop is light and its trilinear coupling $A_t$ is large, it induces 
non-decoupling
effects on electroweak observables (such as the $W$ boson mass $M_W$, 
the effective weak mixing angle $\sin^2\theta_{\rm eff}$, and 
the leptonic decay width of the $Z$ boson $\Gamma_l$, etc). We have 
estimated the stop-sbottom oblique corrections to $M_W$,
$\sin^2\theta_{\rm eff}$, and $\Gamma_l$ and found
that a large left-right mixing of sbottoms with a light 
sbottom (close to the current experimental mass bound) is preferred 
in order to compensate the effects from 
the stop in the scenarios under consideration. The presence of 
light sbottom does not strongly modify the above 
results,\footnote{If the sbottom sector has an additional 
CP-violating phase, the light sbottom can play an important role in 
the ``CP-odd'' Higgs boson production when $\tan\beta$ is large.}
though it could lead to interesting phenomenology 
at current and future colliders. 
The third one comes from EDMs 
of electron, neutron and mercury. When $A_t$ has a CP-violating phase and 
the stop and Higgs bosons
are relatively light, two-loop diagrams through stops and Higgs boson 
mediation can induce
large contributions to the EDMs~\cite{EDMs}.  The two-loop
contributions to the electron and neutron EDMs 
have been given in Ref.~\cite{EDMs}.  From that
we found those contributions are typically larger than the current
experimental bounds in the parameter space discussed in this letter.
Therefore, if these two-loop contributions are the only contributions to
the EDMs, the possibilities we have discussed above would have been excluded. 
In order to 
avoid the EDM constraints, one can increase the stop and the 
``CP-odd'' Higgs boson masses and still find the same effect discussed above. 
However, the production cross section of ``CP-odd'' Higgs boson
will become smaller (for a larger mass), and hence it will be difficult 
to find the ``CP-odd'' Higgs boson even at the LHC.
In the general MSSM, however, we cannot exclude a possibility
that cancellations happen~\cite{EDM_cancel} 
among many contributions to the EDMs (not only two-loop
contributions induced by stop and Higgs boson but also one-loop contributions
and/or other two-loop contributions to the EDMs) since many other CP-phases
in the first and second generation squarks and sleptons
can contribute largely to the EDMs but very little 
to Higgs boson physics.  Therefore, the
searches for the large enhancement in the ``CP-odd'' Higgs boson production
may provide an important information on the origin of CP-violation,
independently of the EDM searches.
Other possible constraints will come from B- and K-physics, which, however,
depend strongly on the flavor structure in supersymmetry breaking. 
For example, our scenarios with a light stop will not contradict
the $b\rightarrow s \gamma$ data if there is extra flavor violation
in the squark sector.
Therefore, we do not consider the constraints from B- and K-physics in our 
analysis.

In this letter, we have discussed the effect of CP-violating interaction
in the stop
sector on the ``CP-odd'' Higgs boson production via gluon fusion.
We found that the cross section can be enhanced by a factor of 
about 1000 when $A_t$ and $\mu$ are large, especially when
$m_A>2 m_{\tilde{t}_1}$. When the ``CP-odd'' Higgs boson can decay into
a pair of stops, the stop pair production will be an interesting signature
of CP-violation. When the ``CP-odd'' Higgs boson is not 
kinematically allowed to decay into any superparticles, 
the $A\to \gamma \gamma$ and $\tau\tau$ modes
can be important discovery modes at the LHC.
Although in order to avoid the EDM constraints 
one needs some unnatural fine tunings in the EDMs or
needs to make the Higgs boson and the stop heavier, the searches for
the ``CP-odd'' Higgs boson in the CP-violating case will give us an 
important information on the nature of CP-violation.

In the decoupling limit ($\alpha \sim \beta-\pi/2$), 
the interactions of the heavier ``CP-even''
Higgs boson $H^0$ with $\tilde{t}_L$ and $\tilde{t}_R$
take a similar form as those of the ``CP-odd'' Higgs boson $A$.
Therefore, we expect that similar enhancement would also
apply to $H^0$ production when $A_t$ and $\mu$ are large
even in the case without CP violation in the stop 
sector~\cite{future_work}.

We thank A. Belyaev, K. Hagiwara and P. Zerwas for useful comments
and C.-R. Chen for collaboration at the early stages.
This work was supported in part by the U. S. National Science
Foundation under awards PHY-0244919 and PHY-0354838.
C.-P.~Y. is grateful for the hospitality of National Center for
Theoretical Sciences in Taiwan, R.\ O.\ C., where part of this work
was performed.
K.~T. gratefully acknowledges the hospitality of Osaka University, 
University of Tokyo, Tohoku University and Summer Institute 2005 at 
Fujiyoshida, where part of this work was done.


%
\end{document}